\documentclass[sigconf]{acmart}

\setcopyright{none}
\renewcommand\footnotetextcopyrightpermission[1]{}
\AtBeginDocument{%
  }

\usepackage{microtype}
\usepackage{listings}
\usepackage{xspace}
\usepackage{pifont}
\usepackage{tikz}
\usetikzlibrary{shapes.geometric,arrows.meta,positioning,fit,backgrounds,calc}
\newcommand{\cmark}{\ding{51}}
\newcommand{\xmark}{\ding{55}}

\newcommand{\sys}{\textsc{BranchFS}\xspace}
\newcommand{\syscall}{\texttt{branch()}\xspace}

\begin{document}

\title{Fork, Explore, Commit: OS Primitives for Agentic Exploration}

\acmConference[Agentic OS Workshop]{Agentic OS Workshop, ASPLOS 2026}{March 23, 2026}{Pittsburgh, USA}

\author{Cong Wang}
\authornote{Corresponding author.}
\affiliation{%
	\institution{Multikernel Technologies, Inc.}
	\country{USA}
}
\email{cwang@multikernel.io}

\author{Yusheng Zheng}
\affiliation{%
	\institution{University of California, Santa Cruz}
	\country{USA}
}
\email{yzhen165@ucsc.edu}

\begin{abstract}
	AI agents increasingly perform agentic exploration: pursuing multiple solution paths in parallel and committing only the successful one. Because each exploration path may modify files and spawn processes, agents require isolated environments with atomic commit and rollback semantics for both \emph{filesystem state} and \emph{process state}. We introduce the \emph{branch context}, a new OS abstraction that provides: (1)~copy-on-write state isolation with independent filesystem views and process groups, (2)~a structured lifecycle of fork, explore, and commit/abort, (3)~first-commit-wins resolution that automatically invalidates sibling branches, and (4)~nestable contexts for hierarchical exploration. We realize branch contexts in Linux through two complementary components. First, \sys is a FUSE-based filesystem that gives each branch context an isolated copy-on-write workspace, with O(1) creation, atomic commit to the parent, and automatic sibling invalidation, all without root privileges. \sys is open sourced in \url{https://github.com/multikernel/branchfs}, along with a Python integration library, \textsc{BranchContext}~\cite{branchcontext}, that provides ready-to-use agent exploration patterns. Second, \syscall is a proposed Linux syscall that spawns processes into branch contexts with reliable termination, kernel-enforced sibling isolation, and first-commit-wins coordination. Preliminary evaluation of \sys shows sub-350\,$\mu$s branch creation independent of base filesystem size, and modification-proportional commit overhead (under 1\,ms for small changes).
\end{abstract}


\maketitle

\section{Introduction}

Modern AI agents autonomously execute multi-step tasks across domains such as software engineering~\cite{swe-agent,openhands,claudecode}, computer use~\cite{wang_llm_agent_survey}, and OS tuning~\cite{kgent} through a reason-then-act loop~\cite{react}, where each action (shell commands, file edits, package installations) may produce irreversible side effects. Increasingly, agent systems explore multiple solution paths in parallel~\cite{tree-of-thoughts,reflexion,swe-agent}: for example, a software engineering agent may try several candidate fixes simultaneously, committing only the one that passes tests. We term this pattern \emph{agentic exploration}: the autonomous, parallel pursuit of multiple solution paths with selective commitment of results. It requires isolating two types of state across parallel branches: \emph{filesystem state} (workspace modifications) and \emph{process state} (execution context), with the ability to commit successful branches and discard failed ones.

Existing OS mechanisms do not meet these requirements individually, and composing them is fragile: filesystem branching primitives lack nested branching, commit semantics, and unprivileged operation; process management mechanisms lack reliable branch-aware termination and sibling isolation; and stitching these together in userspace introduces race windows between steps, error-prone cleanup on partial failure, and ordering dependencies that defeat correctness (Section~\ref{sec:motivation}). In practice, current agent frameworks resort to ad hoc solutions such as git stashing, temporary directories, or container clones~\cite{langchain,autogpt}, which incur significant overhead and cannot capture all filesystem modifications.

We introduce the \emph{branch context}, a new OS abstraction that captures these requirements (Section~\ref{sec:branch-contexts}). A branch context is an isolated execution environment, with a copy-on-write filesystem view and a process group, that follows a fork/explore/commit lifecycle with first-commit-wins resolution and support for nesting. We realize branch contexts in Linux through two complementary components:

\begin{enumerate}
	\item \textbf{\sys}: A FUSE-based~\cite{fuse} filesystem that gives each branch context an isolated copy-on-write workspace, with O(1) creation, atomic commit to the parent, and automatic sibling invalidation, all without root privileges.

	\item \textbf{\syscall}: A proposed Linux syscall that spawns processes into branch contexts with reliable termination, kernel-enforced sibling isolation, and first-commit-wins coordination.
\end{enumerate}

\noindent Preliminary evaluation shows sub-350\,$\mu$s branch creation and under 1\,ms commit for small changes (Section~\ref{sec:evaluation}).

We make the following contributions:
\begin{itemize}
	\item The \emph{branch context}, a new OS abstraction for agentic exploration with copy-on-write isolation, fork/explore/commit lifecycle, first-commit-wins resolution, and nesting (Section~\ref{sec:branch-contexts}).
	\item Two complementary Linux implementations: \sys, a FUSE-based branching filesystem (Section~\ref{sec:branchfs}); and \syscall, a proposed syscall for process coordination (Section~\ref{sec:branch-syscall}).
	\item A preliminary evaluation of \sys (Section~\ref{sec:evaluation}).
\end{itemize}

\section{Motivation}
\label{sec:motivation}

\subsection{AI Agent Execution Patterns}

Several agent patterns amplify the side-effect challenge introduced above by exploring multiple execution paths: Tree/Graph-of-Thoughts reasoning~\cite{tree-of-thoughts,graph-of-thoughts} searches over candidate solutions in parallel, Reflexion~\cite{reflexion} retries tasks after self-critique requiring rollback of prior side effects, and software engineering agents such as SWE-agent~\cite{swe-agent} and OpenHands~\cite{openhands} attempt multiple fixes each producing substantial filesystem changes. These patterns share a common need for \emph{lightweight, isolated workspaces}. Current frameworks use ad hoc solutions: LangChain~\cite{langchain} and AutoGPT~\cite{autogpt} rely on git stashing, temporary directories, or container clones, while Claude Code~\cite{claudecode} provides per-file snapshots that cannot capture changes from shell commands (e.g., \texttt{npm install}) and do not support parallel branching. None adequately meets these needs.

\subsection{Requirements of Agentic Exploration}

The agent patterns above impose six requirements on the execution environment:

\paragraph{R1: Isolated parallel execution.}
Because multiple exploration paths run simultaneously and may modify the same files (e.g., editing \texttt{main.py} or running \texttt{npm install}), each path needs its own isolated view of filesystem and process state. Without isolation, concurrent modifications corrupt shared state.

\paragraph{R2: Atomic commit with single-winner resolution.}
When one path succeeds (e.g., tests pass), its changes must be applied atomically to the shared workspace. All sibling paths must be automatically invalidated, since they now operate on stale state and could silently corrupt the committed result. Failed paths must be discardable with no side effects.

\paragraph{R3: Hierarchical nesting.}
Tree-of-Thoughts and similar patterns explore hierarchically: an agent tries three strategies, each of which may try sub-variants. The execution environment must support nested branches, where each level commits to its immediate parent.

\paragraph{R4: Complete filesystem coverage.}
Agent actions go beyond editing source files: they run shell commands that create build artifacts, install packages, and modify dotfiles. The isolation mechanism must capture \emph{all} filesystem modifications, not just tracked files.

\paragraph{R5: Lightweight, unprivileged, and portable.}
Agents create and discard branches frequently (e.g., per LLM reasoning step). Branch creation must be sub-millisecond, require no root privileges, and work portably across underlying filesystems (ext4, XFS, NFS, etc.) since agents run in diverse environments. This rules out heavyweight mechanisms such as VM snapshots, privileged container runtimes, or filesystem-specific solutions.

\paragraph{R6: Process coordination.}
Agents spawn processes (compilers, test runners, package managers) within each path. On commit or abort, all processes in a branch must be reliably terminated; sibling branches must be isolated from each other to prevent cross-path interference (e.g., one branch's test runner killing another's).

\subsection{Limitations of Existing Mechanisms}

No existing OS mechanism satisfies these requirements (Tables~\ref{tab:comparison} and~\ref{tab:process-comparison}).

\paragraph{Filesystem branching.}
Union filesystems such as UnionFS~\cite{unionfs} and its successor OverlayFS~\cite{overlayfs} provide a single writable upper directory with no native commit-to-parent or sibling invalidation, and mounting requires root privileges. Btrfs~\cite{btrfs} and ZFS~\cite{zfs} support nested subvolumes and clones but are each tied to a single filesystem type and lack native commit-to-parent. Device-mapper snapshots~\cite{dmsnapshot} require raw block devices with O(depth) read latency.

At coarser granularity, VM snapshot systems~\cite{qcow2,vmware-snapshots,remus,snowflock} have whole-VM overhead; recent serverless snapshot work~\cite{firecracker,faasnap,catalyzer} reduces startup latency but does not address branching semantics; and CRIU~\cite{criu} requires full state serialization, all violating R5.

\paragraph{Process management.}
Process groups and sessions~\cite{apue} allow group termination, but processes can escape via \texttt{setsid()} or \texttt{setpgid()}; manual PID tracking via \texttt{/proc} suffers from race conditions. Cgroups~\cite{cgroups} provide reliable termination but require setup and typically need root privileges (cgroup~v2 delegation via systemd can avoid root but requires prior configuration), and do not prevent signaling between siblings. PID namespaces provide complete isolation but impose PID~1 init overhead. The \texttt{clone()} syscall~\cite{clone} enables fine-grained resource sharing but requires combining multiple mechanisms. None satisfies R6 in full.

\begin{table}[t]
	\centering
	\caption{Feature comparison of filesystem branching mechanisms.}
	\label{tab:comparison}
	\resizebox{\columnwidth}{!}{%
		\begin{tabular}{lcccc}
			\toprule
			Feature              & OverlayFS/UnionFS & Btrfs/ZFS & DM-Snap & \sys   \\
			\midrule
			Portable across FSes & \cmark    & \xmark & \cmark  & \cmark \\
			Nested branches      & \xmark    & \cmark & \xmark  & \cmark \\
			Commit-to-parent     & \xmark    & \xmark & \xmark  & \cmark \\
			Sibling invalidation & \xmark    & \xmark & \xmark  & \cmark \\
			No root privileges   & \xmark    & \xmark & \xmark  & \cmark \\
			\bottomrule
		\end{tabular}}

\end{table}

\begin{table}[t]
	\centering
	\caption{Feature comparison of process management mechanisms for agentic exploration.}
	\label{tab:process-comparison}
	\resizebox{\columnwidth}{!}{%
		\begin{tabular}{lccccc}
			\toprule
			Feature              & pgrp   & session & cgroup & PID ns & \syscall \\
			\midrule
			Reliable termination & \xmark & \xmark  & \cmark & \cmark & \cmark   \\
			No escape possible   & \xmark & \xmark  & \cmark & \cmark & \cmark   \\
			Sibling isolation    & \xmark & \xmark  & \xmark & \cmark & \cmark   \\
			No setup required    & \cmark & \cmark  & \xmark & \xmark & \cmark   \\
			No root privileges   & \cmark & \cmark  & \xmark$^{\dagger}$ & \xmark & \cmark   \\
			No PID~1 complexity  & \cmark & \cmark  & \cmark & \xmark & \cmark   \\
			Exclusive groups     & \xmark & \xmark  & \xmark & \xmark & \cmark   \\
			\bottomrule
		\end{tabular}}

	\smallskip
	{\scriptsize $^{\dagger}$Cgroup~v2 delegation via systemd can avoid root but requires prior configuration.}
\end{table}

\section{Branch Contexts}
\label{sec:branch-contexts}

\begin{figure}[t]
	\centering
	\begin{tikzpicture}[
		node distance=0.4cm and 0.3cm,
		process/.style={rectangle, draw, fill=blue!20, minimum width=1.8cm, minimum height=0.6cm, font=\small},
		syscall/.style={rectangle, draw, fill=orange!30, minimum width=2.5cm, minimum height=0.6cm, font=\small\ttfamily},
		namespace/.style={rectangle, draw, dashed, fill=green!10, minimum width=1.6cm, minimum height=0.8cm, font=\scriptsize},
		branch/.style={rectangle, draw, fill=yellow!20, minimum width=1.6cm, minimum height=0.6cm, font=\scriptsize},
		base/.style={rectangle, draw, fill=gray!20, minimum width=6cm, minimum height=0.6cm, font=\small},
		fuse/.style={rectangle, draw, fill=purple!15, minimum width=6cm, minimum height=0.6cm, font=\small},
		arrow/.style={-{Stealth[length=2mm]}, thick},
		label/.style={font=\scriptsize\itshape}
		]
		\node[process] (parent) {Parent Process};

		\node[syscall, below=0.5cm of parent] (syscall) {branch(N=3)};

		\node[process, below=0.8cm of syscall, xshift=-2.2cm] (child1) {Child 1};
		\node[process, below=0.8cm of syscall] (child2) {Child 2};
		\node[process, below=0.8cm of syscall, xshift=2.2cm] (child3) {Child 3};

		\node[namespace, below=0.3cm of child1] (ns1) {Mount NS};
		\node[namespace, below=0.3cm of child2] (ns2) {Mount NS};
		\node[namespace, below=0.3cm of child3] (ns3) {Mount NS};

		\node[fuse, below=0.8cm of ns2] (fuse) {\textsc{BranchFS} (FUSE)};

		\node[branch, below=0.5cm of fuse, xshift=-2.2cm] (br1) {Branch 1 ($\Delta$)};
		\node[branch, below=0.5cm of fuse] (br2) {Branch 2 ($\Delta$)};
		\node[branch, below=0.5cm of fuse, xshift=2.2cm] (br3) {Branch 3 ($\Delta$)};

		\node[base, below=0.6cm of br2] (basedir) {Base Directory (Original Files)};

		\draw[arrow] (parent) -- (syscall);

		\draw[arrow] (syscall.south) -- ++(0,-0.2) -| (child1.north);
		\draw[arrow] (syscall.south) -- (child2.north);
		\draw[arrow] (syscall.south) -- ++(0,-0.2) -| (child3.north);

		\draw[arrow] (child1) -- (ns1);
		\draw[arrow] (child2) -- (ns2);
		\draw[arrow] (child3) -- (ns3);

		\draw[arrow] (ns1.south) -- (ns1.south |- fuse.north);
		\draw[arrow] (ns2.south) -- (fuse.north);
		\draw[arrow] (ns3.south) -- (ns3.south |- fuse.north);

		\draw[arrow] (fuse.south) -- ++(0,-0.1) -| (br1.north);
		\draw[arrow] (fuse.south) -- (br2.north);
		\draw[arrow] (fuse.south) -- ++(0,-0.1) -| (br3.north);

		\draw[arrow, densely dotted] (br1.south) -- (br1.south |- basedir.north);
		\draw[arrow, densely dotted] (br2.south) -- (basedir.north);
		\draw[arrow, densely dotted] (br3.south) -- (br3.south |- basedir.north);

		\node[label, right=0.1cm of syscall] {creates N children};
		\node[label, right=0.3cm of ns3] {isolated views};
		\node[label, right=0.3cm of br3] {copy-on-write};

	\end{tikzpicture}
	\caption{Architecture overview. \syscall coordinates process creation and mount namespace isolation; \sys provides filesystem branching with copy-on-write semantics. Dotted arrows indicate copy-on-write fallback to base.}
	\label{fig:architecture}
\end{figure}
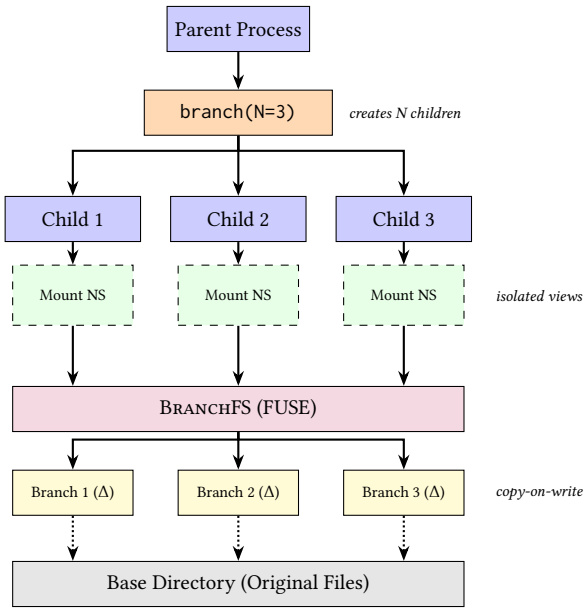

We now introduce the \emph{branch context}, a new OS abstraction that captures the requirements from Section~\ref{sec:motivation}. A branch context is an isolated execution environment, comprising a filesystem view and a process group, that provides copy-on-write state isolation (R1, R4), atomic commit with single-winner resolution (R2), nesting for hierarchical exploration (R3), lightweight unprivileged creation (R5), and coordinated process lifecycle management (R6). Figure~\ref{fig:architecture} illustrates how the two components of our framework (\sys and \syscall) realize this abstraction.

\subsection{Definition}

A branch context~$b_i$ encapsulates: (1)~a filesystem view consisting of the origin's files overlaid with a copy-on-write delta layer~$\Delta_i$, and (2)~a process group whose side effects are confined to~$b_i$. Branch contexts are created in sets of~$N$ siblings from a single \emph{frozen origin}, execute independently, and resolve via atomic commit (promoting one context's changes to the parent) or abort (discarding changes).

Prior work on speculative OS execution introduced related concepts: Tapus et al.~\cite{tapus} proposed ``speculations'' for distributed fault-tolerance, where a single speculative path runs sequentially from a checkpoint and an owner process decides whether to accept results. Branch contexts differ in three ways: they support \emph{parallel} exploration of~$N$ paths simultaneously, use \emph{competitive} first-commit-wins resolution rather than owner-decided acceptance, and target \emph{local} agent workloads rather than distributed system consistency.

\subsection{Lifecycle: Fork, Explore, Commit}

Every branch context follows a three-phase lifecycle:

\paragraph{Fork.}
$N$ branch contexts are created atomically from a frozen origin. Each context~$b_i$ receives its own copy-on-write filesystem layer and process group.

\paragraph{Explore.}
Each branch context executes independently. Filesystem modifications are captured in the per-context delta layer~$\Delta_i$ via copy-on-write; all process side effects are confined to the context.

\paragraph{Commit/Abort.}
When a branch context succeeds (e.g., tests pass), it \emph{commits}: its delta~$\Delta_i$ is applied atomically to the parent. A context may also \emph{abort}, discarding its delta with no effect on the parent or siblings.

\subsection{Core Semantics}

Branch contexts are defined by four core semantic properties:

\begin{enumerate}
	\item \textbf{Frozen origin.} The parent state becomes read-only when branch contexts exist, eliminating merge conflicts by construction.

	\item \textbf{Parallel isolated execution.} All $N$ contexts run simultaneously, not as sequential retries, and are fully isolated from one another: no context can observe or modify another's state. This enables speedup proportional to available parallelism.

	\item \textbf{First-commit-wins.} Any context may commit; the first to do so wins, and all siblings are invalidated. This avoids merge complexity and suits AI agent workloads where the orchestrator cannot predict which branch will succeed.

	\item \textbf{Nested branch contexts.} A branch context may itself fork sub-contexts, forming a tree. Each level commits to its immediate parent, enabling hierarchical exploration (e.g., an agent tries three strategies, each of which tries two sub-variants).
\end{enumerate}

\noindent The following sections describe how we realize branch contexts: \sys (Section~\ref{sec:branchfs}) implements the filesystem dimension, and \syscall (Section~\ref{sec:branch-syscall}) implements the process coordination dimension.

\section{\sys Design and Implementation}
\label{sec:branchfs}

\sys is a FUSE-based filesystem that realizes the filesystem dimension of branch contexts.

\subsection{File-Level Copy-on-Write}

When a file is first modified on a branch, \sys copies the entire file from the base directory (or nearest ancestor branch that modified it) into the branch's delta layer (implemented as a per-branch directory on disk). Subsequent reads and writes to that file are served entirely from the delta copy. Unmodified files are resolved by walking up the branch chain to the base directory.

This file-level granularity is a deliberate trade-off: it is coarser than block-level copy-on-write but simpler to implement correctly atop FUSE, and the overhead of copying whole files is acceptable for the small-to-medium files typical of agent workloads.

\subsection{Branch Chain and Resolution}

Branches form a chain from the current branch back through ancestor branches to the base directory. File lookups traverse this chain:

\begin{enumerate}
	\item Check the current branch's delta layer.
	\item If not found, check each ancestor branch in order.
	\item If not found in any branch, serve from the base directory.
	\item If a \emph{tombstone marker} is encountered at any level, the file is reported as deleted.
\end{enumerate}

Tombstone markers are lightweight sentinel files that record deletions, ensuring deleted files do not ``reappear'' from the base directory.

\subsection{Commit and Abort Semantics}

\paragraph{Commit.}
A commit applies a branch's changes atomically to its parent:
(1)~Modified files and tombstones are collected from the committing branch's delta layer.
(2)~Deletions (tombstones) are applied first.
(3)~Modified files are copied from the committing branch's delta layer into the parent's delta layer.
(4)~The parent's epoch counter is incremented, invalidating all sibling branches.
(5)~Memory-mapped regions from invalidated branches trigger \texttt{SIGBUS}.

\paragraph{Abort.}
An abort discards the branch's delta layer without modifying the parent, incurring negligible cost. Sibling branches remain valid, and memory-mapped regions from the aborted branch trigger \texttt{SIGBUS}.

\subsection{Mount-Based Isolation and Interface}

\sys can be used standalone, without the \syscall kernel extension. In standalone mode, branches are created, committed, and aborted via a CLI tool (\texttt{branchfs create/\allowbreak commit/\allowbreak abort}) or control files (\texttt{.branchfs\_ctl}) within the filesystem, requiring no kernel modifications. All mounts of the same \sys daemon share a single branch namespace, with each branch accessible as a virtual directory via \texttt{@branch} paths (e.g., \texttt{/mnt/workspace/@feature-a/}). Per-branch isolation for multi-agent workflows is achieved by placing each agent in its own mount namespace with only its \texttt{@branch} path bind-mounted as the workspace root, preventing agents from accessing each other's branches.

When used with \syscall (Section~\ref{sec:branch-syscall}), branch management is instead driven by the kernel: \syscall issues \texttt{FS\_IOC\_BRANCH\_*} ioctls to create, commit, and abort branches, and handles mount namespace setup atomically.

\subsection{Agent Integration Library}
\label{sec:agent-integration}

To bridge \sys with agent frameworks, we provide \textsc{BranchContext}~\cite{branchcontext}, a Python library that wraps \sys's branching primitives into high-level exploration patterns. At its core, a \texttt{Workspace} context manager mounts \sys and exposes \texttt{Branch} objects with automatic commit-on-success and abort-on-failure semantics. On top of this, the library provides seven composable patterns that directly realize the agent strategies described in Section~\ref{sec:motivation}: \emph{Speculate} races $N$ candidates and commits the first success; \emph{BestOfN} runs $N$ candidates and commits the highest-scoring one; \emph{Reflexion} retries sequentially, feeding failure feedback into the next attempt; \emph{TreeOfThoughts} explores hierarchical strategy trees with nested branches; \emph{BeamSearch} keeps the top-$K$ branches alive at each depth level; \emph{Tournament} performs pairwise elimination via a judge function; and \emph{Cascaded} starts with one attempt and adaptively fans out on failure. Each pattern manages branch lifecycle (creation, commit, abort) and process isolation (via \texttt{fork(2)} with optional timeout and resource limits) internally, so agent code only supplies the per-branch task logic.

\section{The \syscall Syscall}
\label{sec:branch-syscall}

While \sys provides filesystem branching, agents must still coordinate processes across branches. We propose \syscall, a Linux syscall that provides this coordination with capabilities that are difficult or impossible to achieve purely in userspace: reliable process termination (no escape via \texttt{setsid()}), kernel-enforced sibling isolation, memory branching via page tables, and PID preservation on commit.

\subsection{Interface}

The \syscall interface is designed around three principles: \emph{simplicity}, exposing only three operations (create, commit, abort) that mirror the branching lifecycle; \emph{flexibility}, allowing callers to select exactly which resources to branch via composable flags; and \emph{extensibility}, using the \texttt{bpf(2)}-style multiplexed syscall pattern so new operations and flags can be added without changing the syscall signature.

The syscall takes an operation code and a union of per-operation argument structures (Listing~\ref{lst:interface}).

\begin{lstlisting}[caption={The \texttt{branch()} syscall interface.},label={lst:interface},float=tp]
/* Operations */
#define BR_CREATE    1  /* Create branch(es) */
#define BR_COMMIT    2  /* Commit to parent */
#define BR_ABORT     3  /* Discard branch */

/* Flags for BR_CREATE */
#define BR_FS        (1 << 0)  /* Branch filesystem */
#define BR_MEMORY    (1 << 1)  /* Branch memory */
#define BR_ISOLATE   (1 << 2)  /* Sibling isolation */
#define BR_CLOSE_FDS (1 << 3)  /* Close all fds */

/* Generic ioctls for branching filesystems */
#define FS_IOC_BRANCH_CREATE  _IO('b', 0)
#define FS_IOC_BRANCH_COMMIT  _IO('b', 1)
#define FS_IOC_BRANCH_ABORT   _IO('b', 2)

union branch_attr {
  struct {  /* BR_CREATE */
    __u32 flags;
    __s32 mount_fd;     /* workspace fd */
    __u32 n_branches;
    __u64 child_pids;   /* pid_t __user *, output */
  } create;
  struct {  /* BR_COMMIT */
    __u32 flags;  /* reserved */
  } commit;
  struct {  /* BR_ABORT */
    __u32 flags;  /* reserved */
  } abort;
};

long branch(int op, union branch_attr *attr,
            size_t size);
\end{lstlisting}

Table~\ref{tab:justification} summarizes why kernel support is necessary for the core capabilities. Beyond individual capabilities, the fundamental argument for a syscall is \emph{atomic composition}: setting up cgroups, PID namespaces, mount namespaces, filesystem branches, and signal barriers in userspace is a multi-step process with race windows between steps (e.g., a process can fork between cgroup creation and migration), fragile cleanup on partial failure, and error-prone ordering dependencies. \syscall composes all of these atomically in a single call with kernel-side cleanup on failure, following the same rationale that motivated \texttt{clone()} over manual \texttt{fork()} + \texttt{unshare()} sequences.

\begin{table}[t]
	\centering
	\caption{Justification for \syscall. Strong justifications require kernel support; weak justifications are conveniences.}
	\label{tab:justification}
	\begin{tabular}{lll}
		\toprule
		Capability            & Userspace?      & Justification   \\
		\midrule
		Atomic composition    & Impossible      & \textbf{Strong} \\
		Memory branching      & Impossible      & \textbf{Strong} \\
		Process termination   & Cgroups only    & \textbf{Strong} \\
		Sibling isolation     & PID ns only     & \textbf{Strong} \\
		Atomic mount setup    & User ns trick   & Weak            \\
		Parent read-only      & mprotect + FUSE & Weak            \\
		First-commit-wins     & Atomics         & Weak            \\
		Multi-branch create   & Fork loop       & Weak            \\
		\bottomrule
	\end{tabular}
\end{table}

\subsection{Operations and Flags}

\paragraph{BR\_CREATE.}
Creates \texttt{n\_branches} child processes, each in its own branch. The kernel issues \texttt{FS\_IOC\_BRANCH\_CREATE} on \texttt{mount\_fd} once per child to create filesystem branches, then atomically forks each child into a new mount namespace and bind-mounts the resulting branch fd over \texttt{mount\_fd} using the new mount API (\texttt{open\_tree} + \texttt{move\_mount}). Returns 0 to the parent (with \texttt{child\_pids} filled) and $1..N$ to each child (branch index). All branches created in a single call form an \emph{exclusive group}: only one can commit successfully; others receive \texttt{-ESTALE}. Four flags control branch scope: \texttt{BR\_FS} (required) gives each child a new mount namespace with its branch mounted as the workspace; \texttt{BR\_MEMORY} additionally branches memory via page table copy-on-write, where writes fault into child-owned pages and the parent's pages become read-only; \texttt{BR\_ISOLATE} adds kernel-enforced signal/ptrace barriers between siblings via \texttt{branch\_id} checks; \texttt{BR\_CLOSE\_FDS} closes inherited file descriptors so children re-open files in their branch context.

\paragraph{BR\_COMMIT.}
The kernel first wins the exclusive group race, then issues \texttt{FS\_IOC\_BRANCH\_COMMIT} on the branch fd to apply filesystem changes, commits memory if \texttt{BR\_MEMORY} was set, and terminates siblings (which receive \texttt{-ESTALE} on their next commit attempt). The committing child then optionally replaces the parent (analogous to \texttt{execve()}), taking over its PID so the caller sees a single continuous process.

\paragraph{BR\_ABORT.}
The kernel issues \texttt{FS\_IOC\_BRANCH\_ABORT} on the branch fd to discard filesystem changes, then terminates the branch process. If all branches abort, the parent resumes.

\paragraph{Nested Branches.}
Branches can create sub-branches by calling \syscall again. A commit replaces only the immediate parent; further propagation requires separate commits.

\subsection{Usage Example}

Listing~\ref{lst:usage} shows the typical pattern for an AI agent exploring three approaches in parallel. The first branch to succeed commits its filesystem changes to the parent; siblings receive \texttt{-ESTALE} and exit.

\begin{lstlisting}[caption={Parallel exploration with first-commit-wins.},label={lst:usage},numbers=left,xleftmargin=2em,float=tp]
int mnt = open("/mnt/workspace", O_PATH);
pid_t pids[3];
union branch_attr attr = {
  .create = { .flags = BR_FS,
               .mount_fd = mnt,
               .n_branches = 3,
               .child_pids = (uintptr_t)pids }
};
int idx = branch(BR_CREATE, &attr, sizeof(attr));
if (idx == 0) {
  /* Parent: wait for winner */
  while (wait(NULL) > 0);
  /* Winner's changes now visible in base */
} else {
  /* Child: idx is 1, 2, or 3 */
  bool ok = try_approach(idx);
  if (ok) {
    union branch_attr ca = {.commit = {0}};
    int r = branch(BR_COMMIT, &ca, sizeof(ca));
    if (r == -ESTALE) _exit(1); /* lost */
    /* r == 0: FS + process committed atomically */
  } else {
    union branch_attr aa = {.abort = {0}};
    branch(BR_ABORT, &aa, sizeof(aa));
    /* does not return */
  }
}
\end{lstlisting}

\subsection{Parent and Child Semantics}

\paragraph{Parent Read-Only.}
While branches exist, the parent becomes read-only for both filesystem (\sys denies writes) and memory (pages marked read-only, writes return \texttt{-EAGAIN}). This eliminates conflicts: the parent spawns branches and waits; on commit, the winning branch replaces it.

\paragraph{Fork and Threads.}
\texttt{fork()} within a branch is permitted: the child inherits the mount namespace (filesystem changes captured in delta) but has a separate address space (memory not part of branch state). Threads via \texttt{clone(CLONE\_VM)} share both mount namespace and address space, so their changes are part of the branch. On commit or abort, all descendants are terminated; the kernel tracks them for reliable cleanup.

\subsection{Filesystem-Agnostic Integration}

\syscall integrates with branching filesystems entirely through three generic ioctls (\texttt{FS\_IOC\_BRANCH\_CREATE}, \texttt{\_COMMIT}, \texttt{\_ABORT}), following the pattern of existing generic ioctls such as \texttt{FICLONE} and \texttt{FIEMAP}. For FUSE-based filesystems like \sys, ioctls are forwarded to the daemon via \texttt{FUSE\_IOCTL}; for kernel-native filesystems, they are handled directly. Crucially, \syscall itself contains no filesystem-specific code: adding a new branching filesystem requires only implementing the three ioctls, with no changes to \syscall or the VFS layer.

\section{Preliminary Evaluation}
\label{sec:evaluation}

We evaluate \sys, our implemented FUSE filesystem. The \syscall design is not yet implemented.

\paragraph{Experimental Setup.}
All experiments run on a machine with an AMD Ryzen 5 5500U CPU (6 cores, 12 threads), 8\,GB DDR4 RAM, and a 240\,GB NVMe SSD (WDC WDS240G2G0C), running Pop!\_OS 22.04 with Linux kernel 6.17. \sys is implemented in approximately 3,400 lines of Rust using the \texttt{fuser} library (FUSE~3 protocol). Each measurement is the median of 10 trials.

\paragraph{Branch Creation.}
Branch creation allocates only a new delta layer directory, independent of base filesystem size. We measure internal operation time (excluding CLI overhead) by instrumenting the daemon. Table~\ref{tab:branch-creation} shows that creation latency remains under 350\,$\mu$s, confirming O(1) cost.

\begin{table}[t]
	\centering
	\caption{Branch creation latency vs.\ base directory size.}
	\label{tab:branch-creation}
	\begin{tabular}{rr}
		\toprule
		Base Size (files) & Creation Latency ($\mu$s) \\
		\midrule
		100               & 292                       \\
		1,000             & 317                       \\
		10,000            & 310                       \\
		\bottomrule
	\end{tabular}
\end{table}

\paragraph{Commit and Abort.}
Commit overhead is proportional to modified data volume, not total filesystem size. For small modifications, commit completes in under 1\,ms. Abort is similarly fast since it only removes the delta layer without copying files to the parent. Table~\ref{tab:commit} shows commit and abort latencies for various modification sizes.

\begin{table}[t]
	\centering
	\caption{Commit and abort latency vs.\ modification size.}
	\label{tab:commit}
	\begin{tabular}{rrr}
		\toprule
		Modification Size & Commit ($\mu$s) & Abort ($\mu$s) \\
		\midrule
		1 KB              & 317             & 315            \\
		100 KB            & 514             & 365            \\
		1 MB              & 2,100           & 890            \\
		\bottomrule
	\end{tabular}
\end{table}

\paragraph{I/O Throughput.}
We measure sequential read and write throughput on a 50\,MB file (64\,KB blocks) in three configurations: native filesystem, \sys with regular FUSE, and \sys with FUSE passthrough. Table~\ref{tab:throughput} summarizes the results.

\begin{table}[t]
	\centering
	\caption{Sequential I/O throughput comparison (MB/s).}
	\label{tab:throughput}
	\begin{tabular}{lrr}
		\toprule
		Mode              & Read  & Write \\
		\midrule
		Native filesystem & 8,800 & 576   \\
		\sys (regular FUSE)      & 1,655 & 631   \\
		\sys (passthrough)       & 7,236 & 719   \\
		\bottomrule
	\end{tabular}
\end{table}

With file descriptor caching, regular FUSE mode achieves 1,655\,MB/s read throughput (19\% of native); the gap reflects the FUSE kernel-to-userspace roundtrip. Write throughput slightly exceeds native because \sys treats \texttt{fsync} as a no-op for ephemeral branches (durability is enforced at commit time), avoiding the SSD sync cost. Passthrough mode (\texttt{FOPEN\_PASSTHROUGH}) bypasses the daemon for unmodified files, reaching 7,236\,MB/s read (82\% of native). These numbers suffice for agent workloads dominated by LLM API latency (100\,ms--10\,s).

\section{Related Work}

\paragraph{Isolation Mechanisms.}
Containers~\cite{docker} combine Linux namespaces~\cite{namespaces} with cgroups~\cite{cgroups} for resource isolation, and gVisor~\cite{gvisor} adds kernel-level sandboxing. MVVM~\cite{mvvm} extends this model to cross-platform agent deployment via WebAssembly enclaves. At a finer granularity, lwCs~\cite{lwc} provide independent protection domains within a process for fast context switching, Dune~\cite{dune} uses hardware virtualization for user-level sandboxing, and Capsicum~\cite{capsicum} introduces capability-based sandboxing for Unix. These systems target general-purpose containment or intra-process isolation, whereas branch contexts focus on lightweight copy-on-write branching with coordinated commit/abort semantics \emph{across} parallel exploration paths.

\paragraph{OS-Level Speculation and Transactions.}
Speculator~\cite{speculator} pioneered speculative execution with output gating for distributed file systems, targeting network latency rather than agent exploration. TxOS~\cite{txos} introduced OS-level ACID transactions, but transactions are flat, short-lived, and require deep kernel modifications. In contrast, branch contexts support nesting, persist for arbitrary durations, and operate entirely in userspace. Baumann et al.~\cite{fork-hotos} argue that \texttt{fork()} is no longer a good abstraction, and \syscall shares this motivation, replacing implicit resource inheritance with explicit branching via composable flags.

\section{Discussion and Future Work}

\paragraph{Semantic Extensions.}
Our current prototype isolates only filesystem state. External side effects (network, IPC, device I/O) are not rolled back on abort. A complete solution requires \emph{effect gating}: buffering external actions until commit and discarding them on abort. Agent gateways such as Agentry~\cite{agentry}, which mediate all agent-to-external communication through a central gateway, provide a natural interposition point, and Speculator~\cite{speculator} demonstrated output gating for distributed file systems. Extending these ideas to general-purpose I/O is future work. On the resolution side, \sys currently supports only single-winner semantics and cannot combine results from multiple branches. Future work will explore multi-branch merge, from file-level union of non-overlapping changes, through conflict detection for overlapping files, to semantic merge strategies.

\paragraph{Implementation Roadmap.}
\syscall is currently a design proposal. We plan to prototype it on Linux 6.19, initially supporting \texttt{BR\_FS} and \texttt{BR\_ISOLATE}, with memory branching (\texttt{BR\_MEMORY}) as a follow-on effort due to the page table complexity involved. On the filesystem side, \sys's file-level copy-on-write has known limitations: symlinks targeting absolute paths outside the branch resolve to unbranched files, hardlinks lose their link relationship on copy-on-write, special files (FIFOs, sockets, device nodes) are unsupported in delta layer, and disk space exhaustion causes \texttt{-ENOSPC} errors. These suffice for agent workloads but need addressing for general use.

Current agents (Claude Code, SWE-agent, OpenHands) primarily checkpoint via files rather than process memory, suggesting that \texttt{BR\_FS}-only mode covers most current use cases. The \texttt{BR\_MEMORY} flag provides future capability for agents with significant in-memory state (e.g., persistent Python interpreters or embedding caches). The \textsc{BranchContext} library (Section~\ref{sec:agent-integration}) already provides this integration for Python-based agents, wrapping \sys into ready-to-use exploration patterns so that agent code only supplies per-branch task logic. Agents run unmodified inside their branch's mount namespace; the library manages branch lifecycle and process isolation.

Although we focus on AI agents, the fork/explore/commit abstraction generalizes: with \texttt{n\_branches=1} it provides single-branch try-and-rollback useful for package management (try an upgrade, abort if it breaks) and system configuration tuning.

\section{Conclusion}

Agentic exploration requires isolating both filesystem and process state across parallel execution paths, with atomic commit and rollback semantics. We introduce the \emph{branch context} abstraction to capture these requirements and present two complementary implementations in Linux. \sys is a FUSE-based filesystem that provides lightweight copy-on-write branching with O(1) creation and atomic commit-to-parent, all without root privileges. On top of \sys, the \textsc{BranchContext} Python library provides seven composable exploration patterns (parallel speculation, best-of-N, reflexion, tree-of-thoughts, and others) for direct integration with agent frameworks. \syscall is a proposed Linux syscall for kernel-enforced process coordination with reliable termination and first-commit-wins semantics, integrating with branching filesystems through generic \texttt{FS\_IOC\_BRANCH\_*} ioctls that keep the syscall filesystem-agnostic. Preliminary evaluation shows O(1) branch creation and modification-proportional commit costs.

\bibliographystyle{ACM-Reference-Format}
\bibliography{references}

\end{document}